\documentclass[conference]{IEEEtran}
\IEEEoverridecommandlockouts
\usepackage{cite}
\usepackage{amsmath,amssymb,amsfonts}
\usepackage{algorithmic}
\usepackage{graphicx}
\usepackage{textcomp}
\def\BibTeX{{\rm B\kern-.05em{\sc i\kern-.025em b}\kern-.08em
    T\kern-.1667em\lower.7ex\hbox{E}\kern-.125emX}}
\begin{document}

\title{Music Transcription by Deep Learning \\ 
        with Data and ``Artificial Semantic'' Augmentation\\
\thanks{The work was partially supported by Ukraine-France Collaboration Project (Programme PHC DNIPRO) (http://www.campusfrance.org/fr/dnipro).}
}

\author{\IEEEauthorblockN{Vladyslav Sarnatskyi*, Vadym Ovcharenko, \\Mariia Tkachenko, Sergii Stirenko, Yuri Gordienko}
\IEEEauthorblockA{\textit{National Technical University of Ukraine} \\
\textit{"Igor Sikorsky Kyiv Polytechnic Institute"}\\
Kyiv, Ukraine \\
*v.sarnatskyi-2019@kpi.ua}
\and
\IEEEauthorblockN{Anis Rojbi}
\IEEEauthorblockA{\textit{Laboratory CHArt (Human and Artificial Cognitions,} \\
\textit{University Paris 8,}\\
Paris, France \\
anis.rojbi@univ-paris8.fr}
}

\maketitle

\begin{abstract}
In this progress paper the previous results of the single note recognition by deep learning are presented. The several ways for data augmentation and ``artificial semantic'' augmentation are proposed to enhance efficiency of deep learning approaches for monophonic and polyphonic note recognition by increase of dimensions of training data, their lossless and lossy transformations.
\end{abstract}

\begin{IEEEkeywords}
machine learning, deep learning, note recognition, data augmentation. 
\end{IEEEkeywords}

\section{Introduction}
Conversion of audio files into musical notation (music transcription) is a popular and very difficult problem even for musicians and experts. That is why the available music transcription tools and methods hardly compete with human perception~\cite{gowrishankar2016exhaustive,knees2016introduction}. 
Recently, several solutions for audio search (Shazam, Soundhound, Doreso) were proposed. For example, in 2003, Avery Li-Chun Wang, a chief scientific engineer at Shazam introduced the audio search algorithm~\cite{wang2003industrial}, where a microphone was used to pick up an audio sample. Then a digital summary of the sound was generated as an acoustic fingerprint, i.e. it was broken down into a simple numeric signature, which was unique to each track and then matched to an extensive audio music database. 
These algorithms are known to perform well on recognizing a short audio sample of music that had been broadcast, mixed with heavy ambient noise, subject to reverb and other processing, captured by a poor microphone, subjected to voice codec compression, etc~\cite{weinstein2005query}. 
Now rethinking of the search by sound problem within the context of current machine learning advances produce surprising results and possibly reveal some intricacies of human hearing that are still not understood. 

The main aim of this short progress paper is to present the previous results of the single note recognition and propose the various means to eliminate ``glass ceiling'' effect in recognition of simultaneously sounding notes by several ways, which are under investigation right now. The section \emph{2.Single Note Recognition} gives the very short outline of the current attempts to use deep learning for the single note recognition. The section \emph{3.Data and ``Artificial Semantic'' Augmentation} proposes several ways for data augmentation and ``artificial semantic'' augmentation to enhance efficiency of machine learning approaches in this context by increase of dimensions of training data, their lossless and lossy transformations. The section \emph{4.Discussion and future work} is dedicated to discussion of the results obtained and future work planned.

\section{Single Note Recognition}
\begin{figure}[h]
\centering
        \includegraphics[scale=0.45]{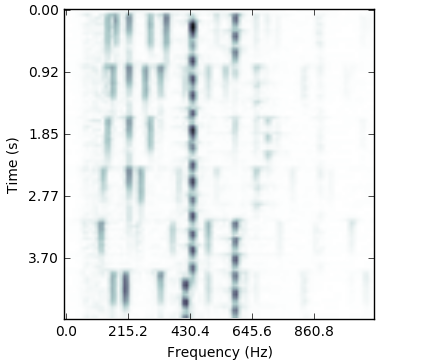}a) \includegraphics[scale=0.45]{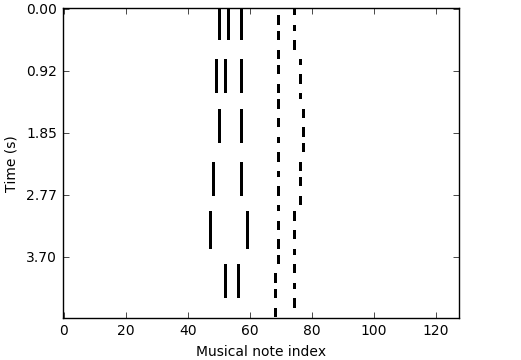}b)
        \caption{Spectrogram of audio file. Dark areas correspond to high amplitude) (a); MNPM of audio file, extracted from MIDI file, corresponding to spectrogram from Fig.~\ref{fig_spectrum}a (white areas correspond to probability 0, black - to 1) (b).}
        \label{fig_spectrum}
\end{figure}

In order to determine presence of certain musical note, frequency characteristics need to be extracted. For this purpose, amplitude time series, gathered from audio file is converted into spectrogram. This process can be done by applying  Discrete Fourier Transform (DFT) to amplitude time series subsequences (Fig.~\ref{fig_spectrum}a).
As far as MIDI files specify many parameters (notation, pitch and velocity, control signals for parameters such as volume, vibrato, audio panning, cues, and clock signals that set and synchronize tempo between multiple devices), the musical note intervals can be defined as a vector $P$ (Fig.~\ref{fig_spectrum}b):
\begin{equation}
	P=(p_i)^{k}_{i=1}, p_i={n_i, b_i, e_i}, i=1,k \label{eq_p_vector}
\end{equation}
where $k$ is the number of note intervals, $n_i$ is the note index of $i$-th
interval, $b_i$ is the timestep index for the $i$-th interval beginning, and $e_i$ is the timestep index for the $i$-th interval end. $P$ can be extracted directly
from MIDI file events.

Then a musical note probability matrix (MNPM) can be defined represented as a rectangular matrix:
\begin{equation}
	M=(m_{ij})^{m,n}_{i=1,j=1} \label{eq_MNPM_matrix}
\end{equation}
where $m_{ij}$ is a probability of $j$-th musical note at $i$-th timestep and it can be calculated from the note intervals $P$ as follows:
\begin{equation}
m_{ij} = \left\{ \begin{array}{rl}
 0 &\mbox{ if $ \not\exists p \in P, p=\{n,b,e\}, n=j, b\le i<e$} \\
 1 &\mbox{ if $ \exists     p \in P, p=\{n,b,e\}, n=j, b\le i<e$}
       \end{array} \right.
	\label{eq_m_P_extract_matrix}
\end{equation}
Thus, the process of converting the audio file into MIDI format can be interpreted as applying model, capable of mapping spectrogram into musical note probability matrix $M$ with further postprocessing.

\begin{figure}[h]
\centering
        \includegraphics[scale=0.32]{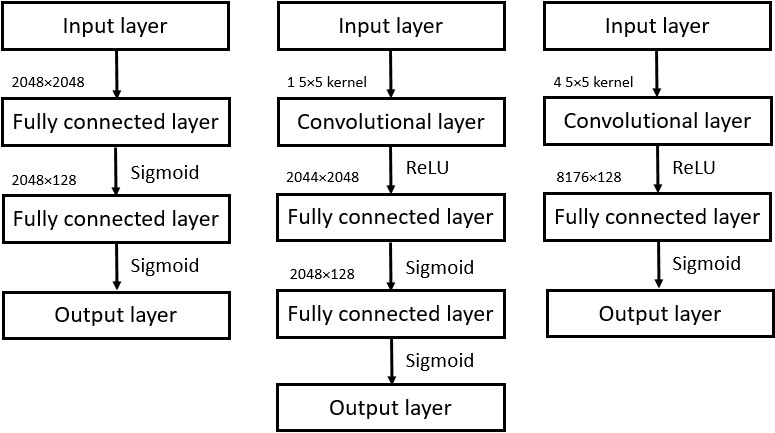}
        \caption{Examples of the neural network architectures used: model A (left) takes a spectrogram as an input; model B (center) and model C (right) take the tensor as an input.}
        \label{fig_NN}
\end{figure}

To map spectrogram into MNPM-matrix $M$ the feedforward artificial neural network was used, but in order to take into account the changes of amplitude with time, $X$ was transformed into tensor. Several artificial neural network architectures were tested (Fig.~\ref{fig_NN}).

\begin{figure}[h]
\centering
        \includegraphics[scale=0.45]{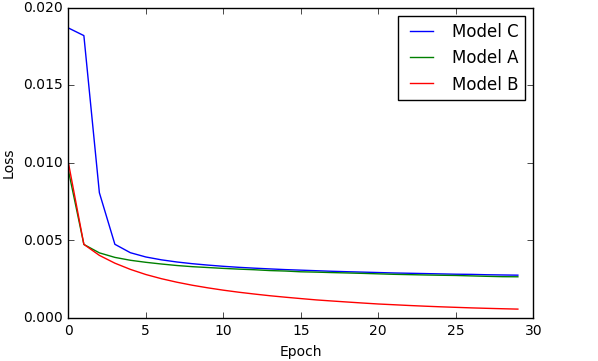}
        \caption{Machine learning rates results for MP3 recordings of piano music.}
        \label{fig_learning}
\end{figure}

Each model produced a 128-vector as an output, that can be interpreted as single row of MNPM. The number 128 comes from MIDI format, which has 128 musical notes. To train each model, audio files spectrograms (i.e. matrix for model A and tensor for models B and C, see Fig.~\ref{fig_NN}) were be fed into it and optimized to minimize loss between model’s output and MNPMs, corresponding to each audio file. The results of this training for the described models are shown in Fig.~\ref{fig_learning}). For training data 95 pairs of MP3 recordings of piano music and corresponding MIDI files were used.

\section{Data and ``Artificial Semantic'' Augmentation}
Despite the fast and steady learning rate the problem of music transcription is aggravated by simultaneously sounding notes (polyphony) from one or more instruments with a complex interaction and appearance of harmonics. Recently, the supervised neural network model for polyphonic piano music transcription was proposed~\cite{sigtia2016end}, which architecture is similar to speech recognition systems and includes an acoustic and language model. The acoustic model is a neural network for estimating the probabilities of pitches in some frame of audio, and the language model is a recurrent neural network for the correlation analysis of pitch combinations over time. Another attempt with a convolutional recurrent neural network (CRNN) shown a strong effectiveness of such hybrid structure for music feature extraction~\cite{choi2017convolutional}. But the other detailed analysis produced a sort of ``glass ceiling'' effect and pessimistic conclusion: the networks can learn combinations of notes, but hardly can recognize the unseen combinations of notes~\cite{kelz2017experimental}. Here, in this progress work paper we propose the various means to eliminate this effect by several data augmentation (like increase of dimensions and lossless transformations of these multidimensional datasets), which are under investigation right now. Data augmentation techniques are very popular now to enrich available datasets by additional data obtained by various ways, mainly, with and without loss of information. Here we propose to use lossless data augmentation techniques that can provide the more data, but create the ``artificial semantics''. 

\subsection{Increase of Dimension}
Increase of dimension is a popular technique, which is currently used for various applications, for example, for discovering cancer precursors from three-dimensional (3D) computer tomography reconstructions than from their original two-dimensional (2D) scans~\cite{ferrante2017slice}. The raw audio signal is 1D can be represented by its 2D spectrogram by analyzing the existing frequencies along with time. The time and the frequency give the dimensions of the spectrogram, and the spectrogram values are represented by the magnitudes of frequencies at some times (Fig.~\ref{fig_spectrum}a). In the context of music transcription it opens opportunities to take into account correlations between notes and include them in the learning process. In this example, 1D time series can be considered as 2D datasets with regard to their power spectrum where additional semantical links can be used for training the neural network. The more promising opportunities for data augmentation by increase of dimension can be elaborated from the additional data channels, for example, in stereophonic, quadrophonic, and other multichannel (like DTS, Dolby Surround, etc.) records. Transition to higher dimensions opens even more data augmentation due to possibility to apply lossless transformations (please, see the next subsection below).

\subsection{Lossless Transformations}
In the context of the multidimensional data inputs (2D, 3D, ...) the effective size of the available dataset can be increased by lossless transformations. For example, 1D vector can be reversed and used as an additional input that effectively duplicate the previous input dataset without loss of data, but with additional ``reverse'' semantics. This added semantics is considered here as ``artificial'', because it is not presented in real life except for exotic cases like backmasking popularised by the Beatles in backward instrumentation on their 1966 album Revolver~\cite{giuliano1999glass}. For a 2D input one can use 8 unique lossless transformations (and for a 3D input one can use 48 unique lossless transformations), actually multiple rotations by 90$^\circ$ and mirror reflections~\cite{conway2016symmetries}. This data augmentation can be considered as an additional source of data from even a single data sample. To increase the actual size of the training dataset we can use all the symmetrical transformations of 2D (and 3D if any) datasets for lossless augmentation. This lossless data augmentation allow us to teach our model to be insensitive to any undesirable distortions of input music data and to be sensitive to the core music pattern itself. The additional promising way is ``crop bootstrapping'', i.e. multiple random cropping the spectrogram along the time axis with random changes of start and end moment.

\subsection{Lossy Transformations}
Usually lossy transformations in machine learning are applied to images and include rotations (not equal to 90$^\circ$), dilations, scale change, addition of noises with various spectral characteristics, etc. Spectrograms cannot be transformed by these transformations (for example, stretched and arbitrary rotated) without negative effect on their provided semantics. For example, any rotation can significantly change the spectrogram and semantics contained in it. But the previous attempts shown that the very low noise do not influence the precision, and the higher noise gives the better precision for single note music transcription, but cannot provide the stable predictions for the more complicated polyphonic data. In addition, the lossy data augmentation is very computationally expensive and it takes additional research in the future to find its advantages. It is especially important in the view of the well-known high sensitivity of deep learning techniques to selection of control parameters like activation function, batch size, dropout ratio, etc.~\cite{kochura2017ysf, kochura2017comparative, kochura2017comparativeperformance}.

\section{Discussion and future work}
The results obtained on single note recognition demonstrated that they cannot be considered seriously for real life applications with polyphonic music. But due to usage of several data augmentation technique they open several opportunities as to the possible ways for the better note recognition. The proposed data augmentation actually includes ``artificial semantics'', which is absent in the original datasets, but it allows to increase the effective size of training data without overfitting. The models trained on the augmented data converge faster, which is explained by the higher volume of the training data, but the results obtained are approximately the same within the limits of error. It should be noted that this approach can be useful for recognition of the symbols, alphabets, and systems used for nonverbal communication~\cite{hamotskyi2017automatized}. This type of communication and automatic recognition methods is especially important for elderly care applications~\cite{gordienko2017augmented}, especially on the basis of the newly available information and communication technologies with multimodal interaction through human-computer interfaces like wearable computing, augmented reality, brain-computing interfaces, etc~\cite{stirenko2017user}.


\begin{thebibliography}{00}
\bibitem{gowrishankar2016exhaustive} Gowrishankar, B. S., and Nagappa U. Bhajantri. ``An exhaustive review of automatic music transcription techniques: Survey of music transcription techniques''. 2016 International Conference on Signal Processing, Communication, Power and Embedded System (SCOPES), IEEE (2016).
\bibitem{knees2016introduction} Knees, Peter, and Markus Schedl. ``Introduction to Music Similarity and Retrieval''. Music Similarity and Retrieval. Springer Berlin Heidelberg, 1-30 (2016).
\bibitem{wang2003industrial} Wang, Avery. ``An Industrial Strength Audio Search Algorithm''. Ismir. Vol. 2003, 7-13 (2003).
\bibitem{weinstein2005query} Weinstein, Eugene. ``Query by humming: a survey''. NYU and Google (2005).
\bibitem{sigtia2016end} Sigtia, Siddharth, Emmanouil Benetos, and Simon Dixon. ``An end-to-end neural network for polyphonic piano music transcription''. IEEE/ACM Transactions on Audio, Speech and Language Processing (TASLP) 24.5, 927-939 (2016).
\bibitem{choi2017convolutional} Choi, K., Fazekas, G., Sandler, M., \& Cho, K. (2017, March). ``Convolutional recurrent neural networks for music classification''.  2017 IEEE International Conference on Acoustics, Speech and Signal Processing (ICASSP), 2392-2396, IEEE(2017).
\bibitem{kelz2017experimental} Kelz, Rainer, and Gerhard Widmer. ``An Experimental Analysis of the Entanglement Problem in Neural-Network-based Music Transcription Systems''. arXiv preprint arXiv:1702.00025 (2017).
\bibitem{ferrante2017slice} Ferrante, Enzo, and Nikos Paragios. ``Slice-to-volume medical image registration: a survey''. Medical Image Analysis 39, 101-123 (2017). 
\bibitem{giuliano1999glass} Giuliano, Geoffrey, and Vrnda Devi. Glass onion: The Beatles in their own words. Da Capo Press (1999). 
\bibitem{conway2016symmetries} Conway, John H., Heidi Burgiel, and Chaim Goodman-Strauss. ``The symmetries of things''. CRC Press (2016). 
\bibitem{kochura2017ysf} Kochura, Yu, et al. ``Comparative Performance Analysis of Neural Networks Architectures on H2O Platform for Various Activation Functions'', IEEE International Young Scientists Forum on Applied Physics and Engineering (YSF-2017) (Lviv, Ukraine),  arXiv preprint arXiv:1707.04940 (2017).
\bibitem{kochura2017comparative} Kochura, Yuriy, et al. ``Comparative Analysis of Open Source Frameworks for Machine Learning with Use Case in Single-Threaded and Multi-Threaded Modes'', arXiv preprint arXiv:1706.02248 (2017).
\bibitem{kochura2017comparativeperformance} Kochura, Yuriy, Sergii Stirenko, and Yuri Gordienko. ``Comparative Performance Analysis of Neural Networks Architectures on H2O Platform for Various Activation Functions'', arXiv preprint arXiv:1707.04940 (2017).
\bibitem{hamotskyi2017automatized} Hamotskyi, Serhii, et al. ``Automatized Generation of Alphabets of Symbols'', arXiv preprint arXiv:1707.04935 (2017).
\bibitem{gordienko2017augmented} Gordienko, Yu, et al. ``Augmented Coaching Ecosystem for Non-obtrusive Adaptive Personalized Elderly Care on the Basis of Cloud-Fog-Dew Computing Paradigm', Proc. 40th International Convention on Information and Communication Technology, Electronics and Microelectronics (MIPRO) Opatija, Croatia (2017),  387-392, ISBN 978-953-233-093-9; arXiv preprint arXiv:1704.04988 (2017).
\bibitem{stirenko2017user} Stirenko, Sergii, et al. ``User-driven Intelligent Interface on the Basis of Multimodal Augmented Reality and Brain-Computer Interaction for People with Functional Disabilities'', arXiv preprint arXiv:1704.05915 (2017).

\end{thebibliography}
\end{document}